\documentclass{PoS}

\title{Pseudo-scalar meson form factors with maximally twisted Wilson fermions 
at $N_f = 2$}

\ShortTitle{Pseudo-scalar meson form factors with maximally twisted Wilson 
fermions at $N_f = 2$}

\author{\speaker{S.~Simula}\\
        Istituto Nazionale di Fisica Nucleare - Sezione di Roma Tre\\
        Via della Vasca Navale 84, I-00146 Roma, Italy\\
        E-mail: \email{simula@roma3.infn.it}}

\author{for the European Twisted Mass Collaboration (ETMC)}

\abstract{We present preliminary results for various electroweak form factors of 
pseudo-scalar mesons using the tree-level improved Symanzik gauge action and 
the maximally twisted mass fermionic action with $N_f = 2$ dynamical flavors.
Our results, obtained for both light and heavy quark masses at a single lattice 
spacing ($a \simeq 0.09$ fm) and at a single lattice volume ($V*T = 24^3*48$), 
exhibit a quite remarkable statistical precision thanks to the use of all-to-all 
quark propagators computed with a stochastic method. 
Moreover very low values of the four-momentum transfer are achieved by making use 
of twisted boundary conditions on the valence quark fields. 
The mass dependence of the pion charge radius is analyzed using Chiral Perturbation 
Theory, obtaining clear evidence of relevant two-loop contributions. 
The universal Isgur-Wise function is computed from heavy-to-heavy electromagnetic 
transitions and its slope in the case of $u(d)$ spectator quarks is found to be 
$\rho_{IW}^2 = 0.77 \pm 0.28$, where the error is statistical only.}

\FullConference{The XXV International Symposium on Lattice Field Theory\\
		 July 30 - August 4 2007\\
		 Regensburg, Germany}

\begin{document}

\section{Introduction}

The European Twisted Mass Collaboration (ETMC) has recently started an intensive, 
systematic program of calculations of three-point correlation functions using the 
large number of gauge configurations produced for three values of the lattice spacing 
and various lattice volumes adopting the tree-level improved Symanzik gauge action 
and the twisted mass fermionic action with $N_f = 2$ dynamical flavors tuned at 
maximal twist (see \cite{LAT07_Carsten}).
The aim is the determination of the electromagnetic (e.m.) and weak semileptonic 
form factors relevant for light and heavy-light mesons as well as for baryons.

In this contribution we present the preliminary results obtained so far for the 
charge form factor of the pion, the universal Isgur-Wise (IW) function, the vector 
and scalar form factors relevant for $K_{\ell 3}$ decays and the $D \to K(\pi)$ 
transition. 

The presently completed runs correspond to three simulated sea-quark masses, $am_{sea} 
= 0.0040,$ $0.0064$ and $0.0100$, at $\beta = 3.9$ corresponding to $a = 0.087(1)$ fm 
($a^{-1} \simeq 2.3$ GeV) \cite{ETMC_PLB07}, and at a single lattice volume ($V*T = 
24^3*48$).
The mass of the spectator valence quark is fixed at the sea-quark mass, while the 
values of the mass of the valence quark struck by the electroweak current are taken 
from the set $\{0.0040, 0.0064, 0.0085, 0.0100, 0.0150,$ $0.022, 0.027, 0.032,$ 
$0.25, 0.32, 0.39,$ $0.46\}$. The first five masses correspond to the ``light'' 
sector and coincide with the values of the sea-quark mass adopted by the ETMC at 
$\beta = 3.9$, the subsequent three are around the ``strange'' quark mass and the 
heaviest four lie in the range from the ``charm'' quark mass $m_c$ to twice $m_c$.
At each value of the sea-quark mass we have computed the two- and three-point 
correlation functions for charged pseudo-scalar mesons, using the standard local 
$\gamma_5$ interpolating fields, on a set of 240 independent gauge configurations, 
separated by 20 consecutive HMC trajectories. 

\begin{figure}[htb]

\parbox{7.5cm}{\centerline{\includegraphics[width=7cm]{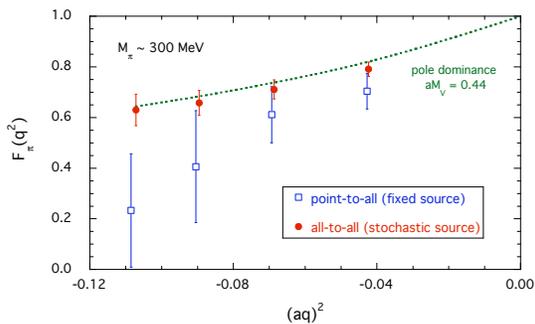}}} \ $~$ \
\parbox{6.5cm}{\vspace{-0.4cm} \caption{Charge form factor of the pion, $F_\pi(q^2)$, 
versus the squared four-momentum transfer $q^2$ in lattice units, calculated at the 
(bare) quark mass $am = 0.0040$ using a subset of 80 ETMC gauge configurations. The 
dotted line is the behavior expected from vector-meson dominance at the simulated 
quark mass. The errors are purely statistical obtained by the jackknife procedure. 
\label{fig:LAT07_fig1}}}

\end{figure}

In order to improve the statistical accuracy we have calculated the two- and 
three-point correlation functions employing all-to-all quark propagators estimated 
through the ``one-end'' stochastic method of Ref.~\cite{one_end_trick}. The advantages 
of such a procedure with respect to the ``standard'' one based on the use of point-to-all 
quark propagators with fixed point sources are clearly illustrated in 
Fig.~\ref{fig:LAT07_fig1}.

Moreover, in order to get rid of the limitations in the minimum value of the spatial 
momentum imposed by periodic boundary conditions, we make use of twisted boundary 
conditions \cite{theta_boundary} on the valence quark fields\footnote{The use of 
different boundary conditions for sea and valence quarks produce finite size 
effects that are exponentially small \cite{SV_PLB05}.}.
We have adopted the Breit frame, where initial and final mesons have opposite spatial 
momenta, because in such a frame for a given value of the squared four-momentum 
transfer $q^2$ the spatial momentum injected to the active quarks is minimized. 
Thus one has
 \begin{eqnarray}
    q^2 = \left[ \sqrt{M_f^2 + \left( \frac{2\pi}{L} \vec{\theta} \right)^2} - 
          \sqrt{M_i^2 + \left( \frac{2\pi}{L} \vec{\theta} \right)^2} \right]^2 - 
          \left( \frac{2\pi}{L} 2 \vec{\theta} \right)^2
    \label{eq:q2}
 \end{eqnarray}
where $M_i$ ($M_f$) is the initial (final) meson mass and $\vec{\theta}$ is a real 
variable. In  our simulations we have chosen $\vec{\theta} = (\tilde{\theta}, 
\tilde{\theta}, \tilde{\theta})$ with $\tilde{\theta}$ ranging from $0.1$ to $1$.

Thanks to the tuning at maximal twist many physical observables at zero momentum are 
automatically $O(a)$-improved \cite{FR_JHEP04}. As for the matrix elements of the 
electroweak current at non-vanishing momenta, the improvement can be realized by a 
suitable averaging of matrix elements with meson momenta of equal magnitude but opposite 
sign \cite{FR_JHEP04}. In the Breit frame this is equivalent to the exchange of initial 
and final quark masses\footnote{Note that the added correlator is almost costless if 
the multisolver algorithm \cite{multisolver} is adopted for the inversion of the Dirac 
equation.}. 
As a byproduct, the matrix elements of the e.m.~current are automatically $O(a)$-improved 
at any momenta in the Breit frame.

\begin{figure}[htb]

\parbox{7.5cm}{\centerline{\includegraphics[width=7cm]{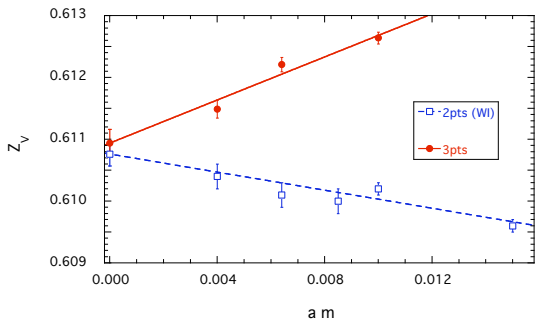}}} \ $~$ \
\parbox{6.5cm}{\vspace{-0.5cm} \caption{Renormalization constant of the local vector 
current $Z_V$ calculated via Eq.~(\protect\ref{eq:ZV}) (full dots) and through the axial 
Ward identity (squares) \cite{LAT07_Petros} for three values of the (bare) quark mass 
$am$. The errors are purely statistical obtained by the jackknife procedure. 
\label{fig:LAT07_fig2}}}

\end{figure}

We employ on the lattice the local vector current which needs to be renormalized. The 
renormalization constant $Z_V$ can be calculated using the matrix element of the time 
component of the (local) vector current between pions at rest. Indeed, since the 
charge form factor of the pion is normalized to unity at $q^2 = 0$, one has
 \begin{eqnarray}
    Z_V \langle \pi^+(\vec{0}) | \frac{2}{3} \overline{u} \gamma^0 u - \frac{1}{3} 
    \overline{d} \gamma^0 d| \pi^+(\vec{0}) \rangle = 2 M_\pi ~ .
    \label{eq:ZV}
 \end{eqnarray}
Another way to obtain $Z_V$ is the use of the axial Ward Identity as carried out
in Ref.~\cite{LAT07_Petros}. The two determinations exhibit a quite remarkable 
statistical precision ($\simeq 0.03 \%$), and they agree very well in the chiral 
limit, as shown in Fig.~\ref{fig:LAT07_fig2}, while at non-vanishing quark masses 
they differ mainly by terms of the order of $a^2 m \Lambda_{QCD}$.

\section{Charge form factor of the pion}

\begin{figure}[htb]

\centerline{\includegraphics[width=14cm]{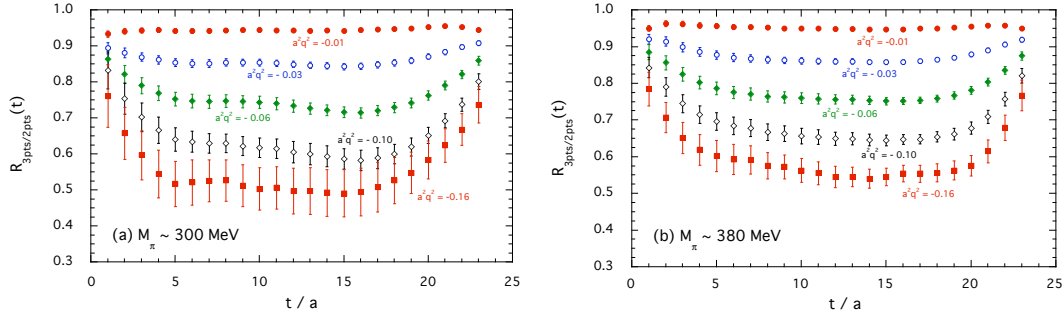}}
\caption{Ratio of three-point to two-point correlation functions, $R_{3pts/2pts}(t)$, 
versus the lattice time for two values of the (bare) quark mass $am = 0.0040$ (a) and 
$0.0060$ (b), corresponding to physical pion mass around $300$ and $380$ MeV, respectively. 
The plateaux of $R_{3pts/2pts}(t)$ provide directly the form factor (\protect\ref{eq:Fpi}).
\label{fig:LAT07_fig3}}

\end{figure}

The charge form factor of the pion, $F_\pi(q^2)$, is directly related to the matrix 
element of the time component of the renormalized (local) e.m. current by
 \begin{eqnarray}
    F_\pi(q^2) = \frac{Z_V}{2E_\pi} \langle \pi^+(\vec{\theta}) | \frac{2}{3} \overline{u} 
    \gamma^0 u - \frac{1}{3} \overline{d} \gamma^0 d| \pi^+(-\vec{\theta}) \rangle ~ .
    \label{eq:Fpi}
 \end{eqnarray}
where $E_\pi = \sqrt{M_\pi^2 + ( 2\pi \vec{\theta} / L)^2}$ and $q^2 = - 4 (2\pi 
\vec{\theta} / L)^2$. Note that the value of $q^2$ is independent of the simulated pion 
mass. 
The matrix element appearing in Eq.~(\ref{eq:Fpi}) can be extracted from a suitable ratio 
of three-point to two-point correlation functions. The quality of the plateaux is 
illustrated in Fig.~\ref{fig:LAT07_fig3}. We remind that in the two-point correlator
the identification of the pion ground state starts already at a time around $t/a = 10$ 
(see \cite{LAT07_Vittorio}).

\begin{figure}[htb]

\centerline{\includegraphics[width=14cm]{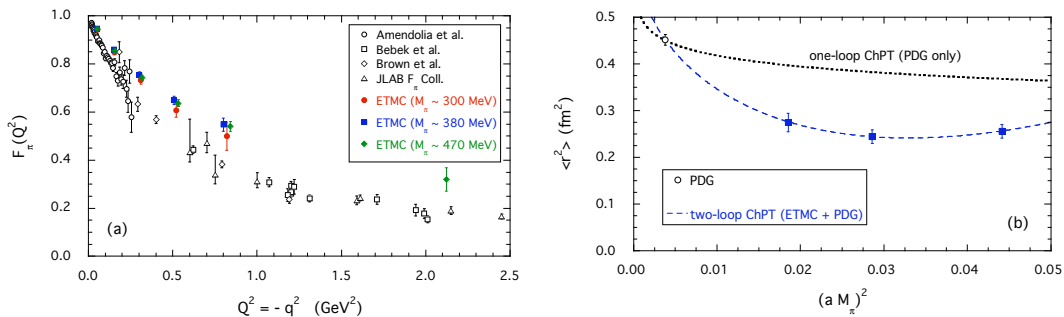}}
\caption{ (a) Charge form factor of the pion $F_\pi(Q^2)$ versus $Q^2 \equiv - q^2$ 
(in physical units). The open markers are the experimental data of Ref.~\cite{pion_data}, 
while the full markers are the ETMC results. The lattice points at $M_\pi \sim 380$ and 
$470$ MeV are slightly shifted in $Q^2$ for a better reading. (b) Squared pion charge 
radius (in physical units) versus the squared pion mass (in lattice units). The 
experimental point is from PDG \cite{PDG}. The dotted and dashed lines are the 
results of ChPT fits at one-loop and two-loops, respectively (see text). 
\label{fig:LAT07_fig4}}

\end{figure}

The results obtained for $F_\pi(Q^2 \equiv -q^2)$ in the unitary setup (i.e., equal 
valence and sea quark masses) are reported in Fig.~\ref{fig:LAT07_fig4}(a) and compared 
with experimental data from Ref.~\cite{pion_data}. It can be seen that: i) the lattice 
results exhibit a remarkable statistical precision; ii) thanks to the use of twisted 
boundary conditions the form factor is precisely determined at values of $Q^2$ as low 
as $0.05$ GeV$^2$; and iii) the lattice results, obtained at pion masses of about $300, 
380$ and $470$ MeV, overestimate the experimental data in the whole range of values of 
$Q^2$.

The $q^2$-dependence of our ETMC results can be very nicely fitted assuming a pole 
behavior. The corresponding values of the pion charge radius are shown in 
Fig.~\ref{fig:LAT07_fig4}(b) and lie well below the experimental value from PDG 
\cite{PDG}. 

The predictions of Chiral Perturbation Theory (ChPT) for the pion charge radius have 
been elaborated at one-loop in Ref.~\cite{GL_84} and at two-loop in Ref.~\cite{BCT_98} 
(in the continuum and infinite volume limits). At one-loop one has
 \begin{eqnarray}
    \langle r^2 \rangle = \frac{2}{(4\pi F)^2} \left[ \mbox{log}(\Lambda_6^2 / M_\pi^2) - 
                          1 \right] ~ ,
    \label{eq:one_loop}
 \end{eqnarray}
where $F$ is the pion decay constant in the chiral limit and $\Lambda_6$ a low-energy 
constant (LEC), while the two-loop formula of Ref.~\cite{BCT_98} can be rewritten as
 \begin{eqnarray}
    \langle r^2 \rangle = \frac{2}{(4\pi F)^2} \left[ \mbox{log}(\Lambda_6^2 / M_\pi^2) - 
                          1 \right]  + A M_\pi^2 + B M_\pi^2 ~ \mbox{log}(M_\pi^2) ,
    \label{eq:two_loop}
 \end{eqnarray}
where $A$ and $B$ depend on various LEC's.
The value of the decay constant $F$ has been determined for our unitary setup in 
Ref.~\cite{ETMC_PLB07} ($aF = 0.0534(6)$ corresponding to $F \simeq 121$ MeV). Thus 
the experimental value of the pion charge radius fix the value of the LEC $\Lambda_6$ 
in the one-loop formula (\ref{eq:one_loop}), namely $\overline{\ell_6} \equiv 
[\mbox{log}(\Lambda_6^2 / M_\pi^2)]_{M_\pi = 139.6 ~ MeV} = 14.4(3)$, as well as its 
mass dependence, as shown in Fig.~\ref{fig:LAT07_fig4}(b) by the dotted line. 
The differences of the one-loop ChPT prediction and the ETMC points represent a clear 
indication of important contributions from higher loops. 

In this preliminary analysis we neglect both finite size and discretization effects, 
which are nevertheless expected to be small, and we use the two-loop formula 
(\ref{eq:two_loop}), containing three free parameters, to fit both the three ETMC 
points and the PDG value, obtaining $\overline{\ell_6} = 17.2 (7)$, as shown by 
the dashed line in Fig.~\ref{fig:LAT07_fig4}(b)). 
It can be seen that the chiral enhancement expected at low pion masses is hardly 
visible at the simulated pion masses. More lattice points, particularly below 
$M_\pi \sim 300$ MeV, are necessary to find out a clear signature of the chiral 
logs. 

We have calculated also the scalar form factor of the pion, limiting ourselves only
to the connected insertion of the scalar density operator. The results will be presented 
elsewhere \cite{ETMC_3pts}. We simply want to point out that the values obtained for 
the scalar radius exhibit the same qualitative features of those discussed above 
for the pion charge radius, including the relevance of two-loop effects in their 
pion mass dependence.

\section{Universal Isgur-Wise function}

The investigation of heavy-to-heavy e.m.~transitions, described by a single form factor
 \begin{eqnarray}
    F_{PS}(q^2) = \frac{Z_V}{2E_{PS}} \langle PS(\vec{\theta}) | \overline{h} 
    \gamma^0 h | PS(-\vec{\theta}) \rangle ~ ,
    \label{eq:IW}
 \end{eqnarray}
allows to determine the IW function $\xi(\omega)$ by performing the infinite heavy-quark 
limit, viz.~$\xi(\omega) = \mbox{lim}_{m_h \to \infty} ~ F_{PS}(q^2)$, where $\omega = 
1 - q^2 / 2 M_{PS}^2$\footnote{With such a definition we disregard the (small) 
perturbative correction that should be removed from Eq.~(\ref{eq:IW}) to get the 
proper renormalization-group invariant definition of $\xi(\omega)$.}. 
Note that the form factor $F_{PS}(q^2)$ is automatically normalized to unity at 
$q^2 = 0$ for any simulated mass because of the conservation of the e.m.~current.

\begin{figure}[htb]

\centerline{\includegraphics[width=14cm]{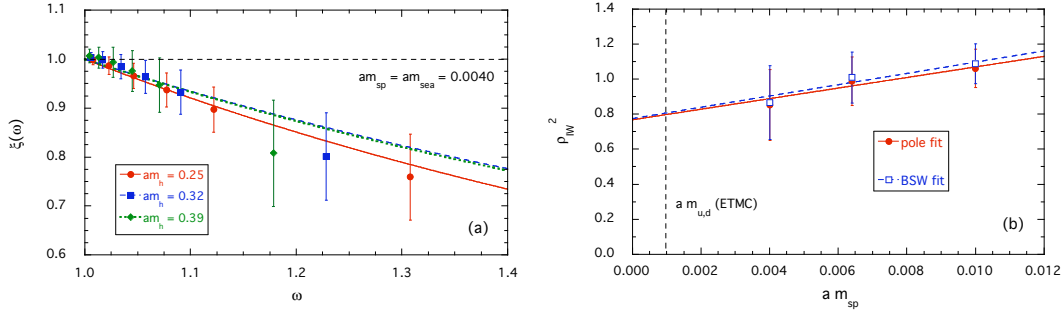}}
\caption{ (a) The IW function calculated at $a m_{sp} = a m_{sea} = 0.0040$ 
for various values of the heavy-quark mass $a m_h$. The various lines are fits 
based on a ``pole'' ans\"atz. (b) Values of the IW slope $\rho_{IW}^2$ obtained 
through ``pole'' (full dots) and BSW (open squares) ans\"atz. The vertical line 
correspond to the value of the (bare) light $u(d)$ quark mass determined in lattice
units in Ref.~\cite{LAT07_Vittorio}. \label{fig:LAT07_fig5}}

\end{figure}

We have calculated $F_{PS}(q^2)$ for various values of the (bare) heavy-quark mass, 
$a m_h$, taking the spectator-quark mass, $a m_{sp}$, to be equal to the sea-quark 
mass $a m_{sea}$. In Fig.~\ref{fig:LAT07_fig5}(a) we have reported the results 
obtained at the lowest sea-quark mass, $a m_{sp} = a m_{sea} = 0.0040$. It can be 
seen that the dependence upon the heavy-quark mass is very mild so that the 
extrapolation to the infinite heavy-quark limit can be safely neglected.

Our results for $\xi(\omega)$ can be nicely fitted using either the pole or the
BSW \cite{BSW_85} ans\"atz. The corresponding values of the IW slope $\rho_{IW}^2 
\equiv - [d \xi(\omega) / d\omega]_{\omega = 1}$ are reported in 
Fig.~\ref{fig:LAT07_fig5}(b). 
A na\"ive linear extrapolation in the spectator quark mass to the (bare) light 
$u(d)$ quark mass, as determined by ETMC in Ref.~\cite{LAT07_Vittorio}, gives 
$\rho_{IW}^2 = 0.77 \pm 0.28$, where the error is statistical only. 
To our knowledge this is the first determination of the IW slope with $N_f = 2$. 
Recently a quite precise value of the IW slope, $\rho_{IW}^2 = 0.89 \pm 0.17$, 
has been obtained ain the quenched approximation in Ref.~\cite{DPT_07}.

\section{$K_{\ell 3}$ decays and heavy-to-light transitions}

\begin{figure}[htb]

\centerline{\includegraphics[width=14cm]{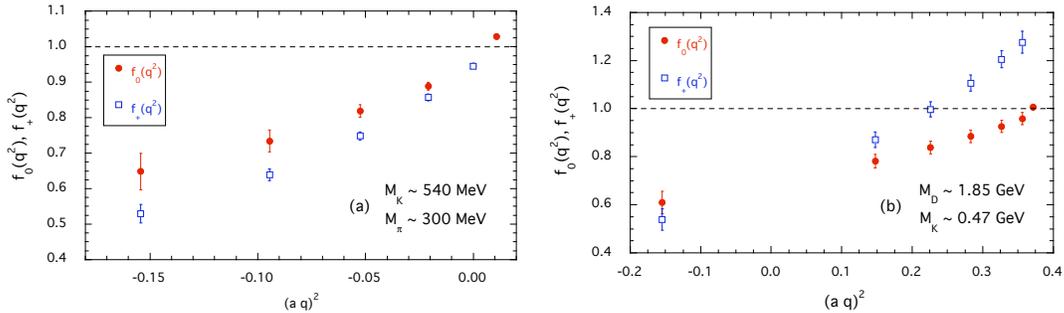}}
\caption{Vector $f_+(q^2)$ and scalar $f_0(q^2)$ form factors of pseudo-scalar mesons 
relevant for $K_{\ell 3}$ decays (a) and for the $D \to K$ transition (b). 
\label{fig:LAT07_fig6}}

\end{figure}

As it is known, the matrix element of the vector weak current between pseudo-scalar 
mesons involve two form factors, the vector $f_+(q^2)$ and the scalar $f_0(q^2)$ 
ones, namely:
 \begin{eqnarray}
    \langle PS_2 | V_\mu^{(weak)} | PS_1 \rangle & = & (p_1 + p_2)_\mu f_+(q^2) +
    (p_1 - p_2)_\mu f_-(q^2) \nonumber \\
    f_0(q^2) & \equiv & f_+(q^2) + \frac{q^2}{M_1^2 - M_2^2} f_-(q^2)
    \label{eq:Kl3}
 \end{eqnarray}
where $q^2 = (p_1 - p_2)^2$.
In this contribution we limit ourselves to illustrate in Fig.~\ref{fig:LAT07_fig6} 
the nice level of statistical precision achieved in the determination of the vector 
and scalar form factors relevant for the cases of $K_{\ell 3}$ decays and of the 
$D \to K$ transition. 

\section{Conclusions}

We have presented preliminary results for the electroweak form factors of light 
and heavy-light pseudo-scalar mesons, obtained at a single lattice spacing ($a 
\simeq 0.09$ fm) and at a single lattice volume ($V*T = 24^3*48$), using the 
tree-level improved Symanzik gauge action and the twisted mass Wilson action 
with $N_f = 2$ dynamical flavors tuned at maximal twist.
The use of all-to-all quark propagators computed with a stochastic method, as well 
as of twisted boundary conditions on the valence quark fields has allowed us to 
achieve both a quite remarkable statical precision and very low values of the 
four-momentum transfer.
We have analyzed the mass dependence of the pion charge radius using Chiral 
Perturbation Theory, obtaining evidence of relevant two-loop contributions.
Results at more values of the sea quark mass, as well as the investigation of volume 
effects and continuum extrapolation, are however required in order to draw definite 
quantitative conclusions.
The universal Isgur-Wise function has been computed from heavy-to-heavy electromagnetic 
transitions and its slope in the case of $u(d)$ spectator quarks has been found to be 
$\rho_{IW}^2 = 0.77 \pm 0.28$, where the error is statistical only.

\end{document}